\documentclass[reprint, amsmath, amssymb, aps, prl, showpacs]{revtex4-1}

\usepackage{graphicx}   
\usepackage{dcolumn}    
\usepackage{bm}     
\usepackage{mathtools}  
\usepackage{caption}
\usepackage{subcaption} 
\usepackage{dsfont}     
\usepackage{soul}       

\newcommand\der[2]{\frac{\text{d}{#1}}{\text{d}{#2}}}
\newcommand\pder[2]{\frac{\partial{#1}}{\partial{#2}}}
\newcommand\inv[1]{\frac{1}{#1}}

\newcommand\mx[1]{\mathbf #1}
\newcommand\R{\textsc{d}}
\renewcommand\L{\textsc{l}}
\newcommand\e{\inv V}

\DeclareRobustCommand{\Chi}{{\mathpalette\irchi\relax}}
\newcommand{\irchi}[2]{\raisebox{\depth}{$#1\chi$}} 

\newcommand\eq[1]{Eq.~(\ref{eq:#1})}

\newcommand\fig[1]{Fig.~\ref{fig:#1}}
\newcommand\Fig[1]{Figure~\ref{fig:#1}}
\newcommand\rx[1]{(\ref{rx:#1})}

\newcounter{defcounter}
\newenvironment{smequation}{%
\addtocounter{equation}{-1}
\refstepcounter{defcounter}

\begin{equation}}
{\end{equation}}

\begin{document}
\title{A noise-induced mechanism for biological homochirality of early life self-replicators}

\author{Farshid Jafarpour}
\author{Tommaso Biancalani}
\author{Nigel Goldenfeld}
\affiliation{
Department of Physics, University of Illinois at Urbana-Champaign,
Loomis Laboratory of Physics, 1110 West Green Street, Urbana, Illinois, 61801-3080,}
\affiliation{Carl R. Woese Institute for Genomic Biology, University of Illinois at Urbana-Champaign,
1206 West Gregory Drive, Urbana, Illinois 61801.}
\date{\today}

\begin{abstract}

The observed single-handedness of biological amino acids and sugars has
long been attributed to autocatalysis. However, the stability of
homochiral states in deterministic autocatalytic systems relies on
cross inhibition of the two chiral states, an unlikely scenario for
early life self-replicators. Here, we present a theory for a stochastic
individual-level model of autocatalysis due to early life
self-replicators. Without chiral inhibition, the racemic state is the
global attractor of the deterministic dynamics, but intrinsic
multiplicative noise stabilizes the homochiral states, in both
well-mixed and spatially-extended systems.  We conclude that
autocatalysis is a viable mechanism for homochirality, without imposing
additional nonlinearities such as chiral inhibition.

\end{abstract}

\pacs{87.23.Kg, 87.18.Tt, 05.40.-a}

\maketitle
\begin{figure*}[t]
    \begin{subfigure}[b]{0.25\textwidth}
            \includegraphics[trim=1cm 0 0 0, width = .90\textwidth]{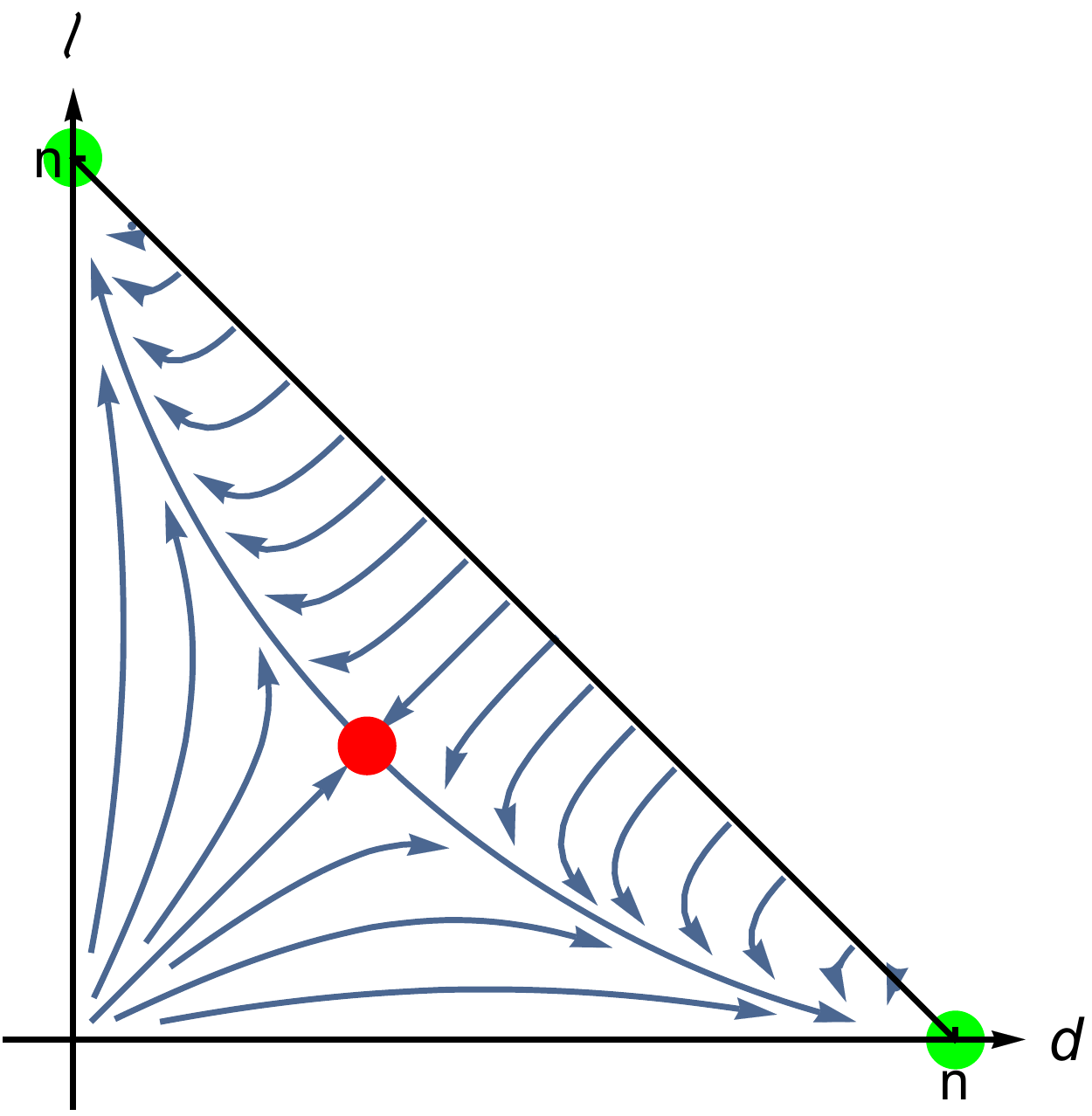}
            \caption{$A+\R\rightarrow 2\R,\hspace{.1in} A+\L\rightarrow 2\L,$\\
                \hspace{.21in}$\R+\L\rightarrow 2A$}
            \label{fig:unstable}
    \end{subfigure}\hspace{0.5in}
    \begin{subfigure}[b]{0.25\textwidth}
        \includegraphics[trim=1cm 0 0 0, width = .90\textwidth]{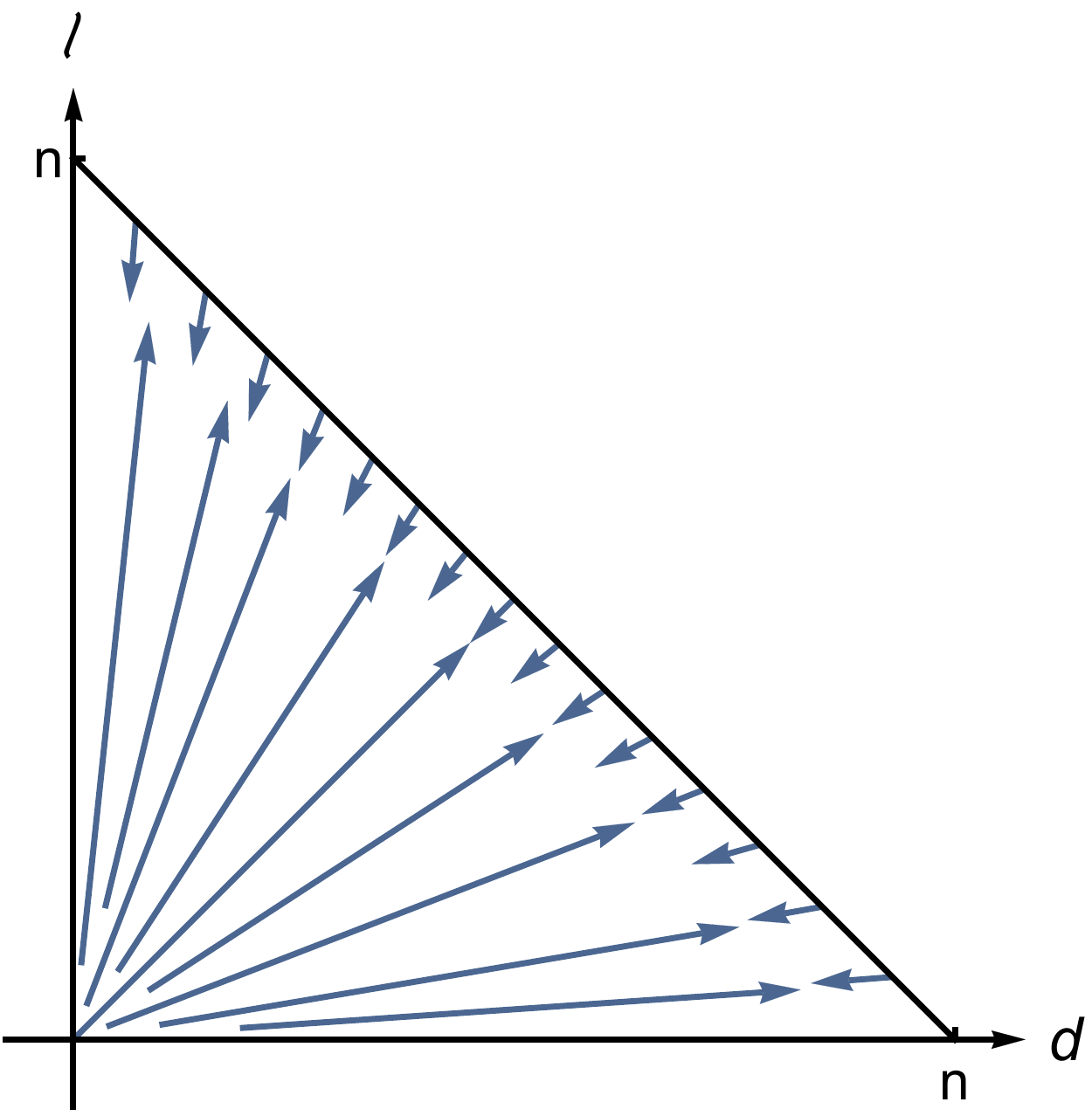}
        \caption{$A+\R\rightarrow 2\R,\hspace{.1in} A+\L\rightarrow 2\L,$\\
            \hspace{.21in}$\R\rightarrow A,\hspace{.1in}  \L\rightarrow A$}
        \label{fig:neutral}
    \end{subfigure}\hspace{0.5in}
    \begin{subfigure}[b]{0.25\textwidth}
        \includegraphics[trim=1cm 0 0 0, width = .90\textwidth]{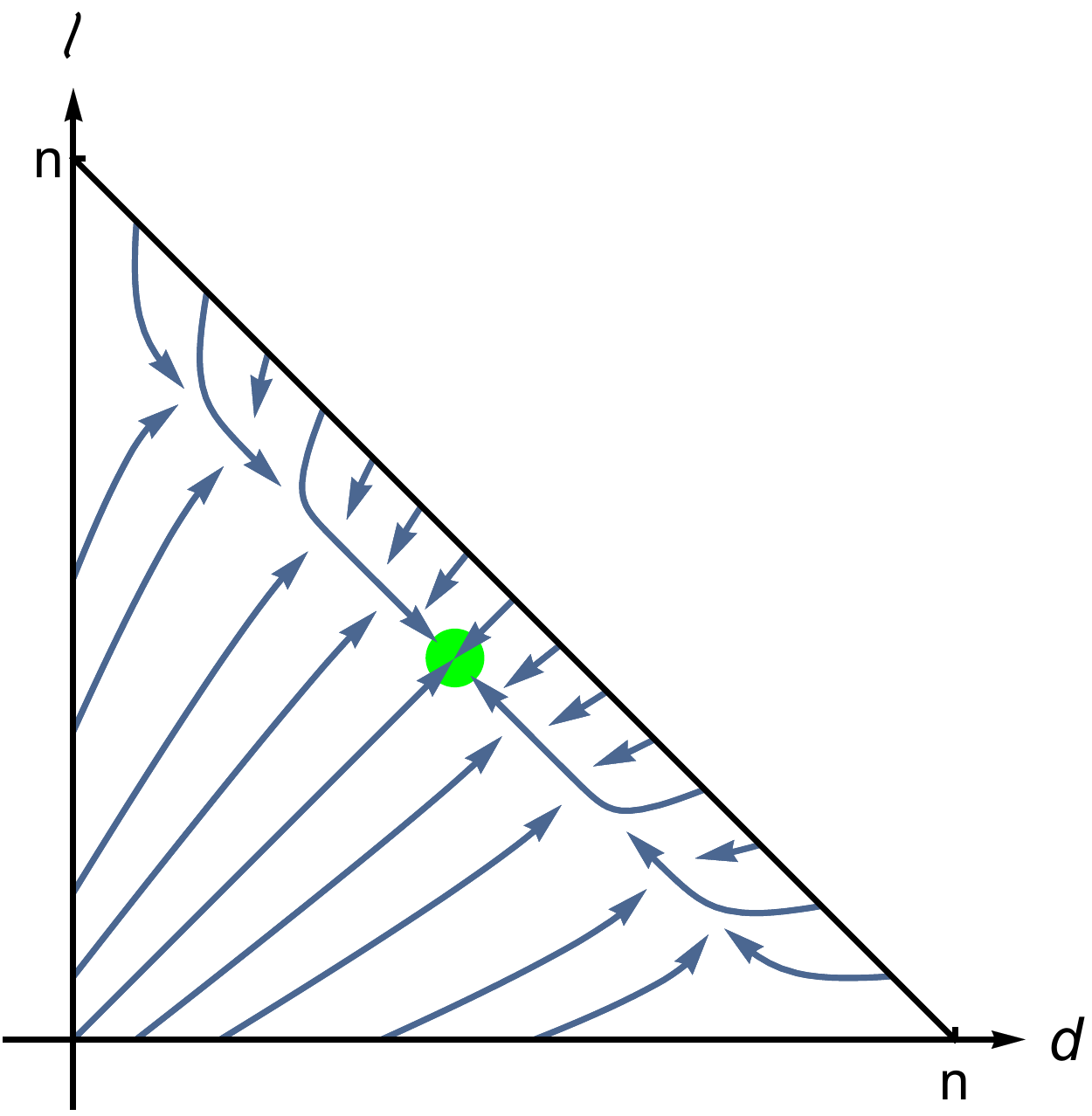}
        \caption{$A+\R\rightarrow 2\R,\hspace{.1in} A+\L\rightarrow 2\L,$\\
            \hspace{.21in}$\R\rightleftharpoons A,\hspace{.1in}  \L\rightleftharpoons A$}
        \label{fig:stable}
    \end{subfigure}
    \caption{(Color online) (a) Phase portrait of Frank\rq{}s model: the
    racemic state is an unstable fixed point (red dot), while the
    homochiral states are stable fixed points (green dots). (b) If chiral
    inhibition is replaced by linear decay reaction, the ratio of $\R$ and
    $\L$ molecules stays constant. (c) Adding even the slightest amount
    of non-autocatalytic production of $\R$ and $\L$ molecules makes
    the racemic state (green dot) the global attractor of the
    dynamics.}
    \label{fig:phase}
\end{figure*}

 One of the very few universal features of biology is
homochirality: every naturally occurring amino acid is left-handed
(\L-chiral) while every sugar is right-handed
(\R-chiral)~\cite{blackmond2010origin, gleiser2012life}. Although such
unexpected broken symmetries are well-known in physics, for example in
the weak interaction, complete biological homochirality still defies
explanation. In 1953, Charles Frank suggested that homochirality could
be a consequence of chemical autocatalysis~\cite{frank1953spontaneous},
frequently presumed to be the mechanism associated with the emergence
of early life self-replicators. Frank introduced a model in which the
$\R$ and $\L$ enantiomers of a chiral molecule are autocatalytically
produced from an achiral molecule $A$ in reactions $A+\R\rightarrow
2\R$ and $A+\L\rightarrow 2\L$, and are consumed in a chiral inhibition
reaction, $\R+\L\rightarrow 2A$ \footnote{In the original model by
Frank, the concentration of the molecules $A$ was kept constant to
reduce the degrees of freedom by one, and the chiral inhibition was
introduced by the reaction $\R+\L\rightarrow \varnothing$. This model
leads to indefinite growth of $\R$ or $\L$ molecules and does not have
a well-defined steady state. To resolve this problem, we let the
concentration of $A$ molecules be variable and replaced this reaction
by $\R+\L\rightarrow 2A$ which conserves the total number of molecules.
This conservation law reduces the number of degrees of freedom by one
again. The mechanism to homochirality in the modified model is the same
as the original model by Frank.}. The state of this system can be
described by the chiral order parameter $\omega$ defined as $\omega
\equiv (d-l)/(d+l)$, where $d$ and $l$ are the concentrations of $\R$
and $\L$. The order parameter $\omega$ is zero at the racemic state,
and $\pm 1$ at the homochiral states. Frank\rq{}s model has three
deterministic fixed points of the dynamics; the racemic state is an
unstable fixed point, and the two homochiral states are stable fixed
points. Starting from almost everywhere in the $\R$-$\L$ plane, the
system converges to one of the homochiral fixed points (\fig
{unstable}).

In the context of biological homochirality, extensions of Frank's idea
have essentially taken two directions. On the one hand, the discovery
of a synthetic chemical system of amino alcohols that amplifies an
initial excess of one of the chiral states~\cite{soai1995asymmetric}
has motivated several autocatalysis-based models
(see~\cite{saito2013colloquium} and references therein). On the other
hand, ribozyme-driven catalyst experiments~\cite{joyce1984chiral}, have
inspired theories based on polymerization and chiral inhibition that
minimize~\cite{sandars2003toy, gleiser2008extended, gleiser2012chiral}
or do not include at all~\cite{plasson2004recycling,
brandenburg2007homochirality} autocatalysis.  In contrast, a recent
experimental realization of RNA replication using a novel ribozyme
shows such efficient autocatalytic behavior that chiral inhibition does
not arise~\cite{sczepanski2014cross}.   Further extensions accounting
for both intrinsic noise~\cite{saito2013colloquium, lente2010role} and
diffusion~\cite{shibata2006diffusion, plasson2006three,
hochberg2006reaction, hochberg2007mirror} build further upon Frank's work.

Regardless of the specific model details, all these models share the
three-fixed-points paradigm of Frank's model, namely that the time
evolution of the chiral order parameter $\omega$ is given by a
deterministic equation of the form~\cite{saito2013colloquium}
\begin{equation} \label{eq:saito}
    \der{\omega}{t} = f(t)\, \omega \left( 1 - \omega^2 \right),
\end{equation}
where the function $f(t)$ is model-dependent. However, the homochiral
states arise from a nonlinearity which is not a property of simple
autocatalysis, but, for instance in the original Frank's model, is due
to chiral inhibition (see~\fig{neutral}). The
sole exception to the three-fixed-points model in a variation of Frank\rq{}s
model is the work of Lente \cite {lente2004homogeneous}, where purely stochastic
chiral symmetry breaking occurs, although chiral symmetry breaking is
only partial, with $\omega\neq 0$ but $|\omega | < 1$.


The purpose of this Letter is to show that efficient early-life
self-replicators can exhibit universal homochirality, through a
stochastic treatment of Frank's model {\em without \/} requiring
nonlinearities such as chiral inhibition. In our stochastic treatment,
the homochiral states arise not as fixed points of deterministic
dynamics, but instead are states where the effects of chemical number
fluctuations (i.e. the multiplicative noise~\cite{Gardiner2009}) are
minimized. The mathematical mechanism proposed here~\cite
{togashi2001transitions, artyomov2007purely, russell2011noise,
biancalani2014noise} is intrinsically different from that of the class
of models summarized by \eq{saito}. In the following, we propose a
model which we analytically solve for the spatially uniform case and
the case of two well-mixed patches coupled by diffusion. We then show,
using numerical simulations, that the results persist in a
one-dimensional spatially-extended system. We conclude that
autocatalysis alone can in principle account for universal
homochirality in biological systems.\\

\noindent {\it Stochastic model for well-mixed system:-} Motivated in
part by the experimental demonstration of autocatalysis without chiral
inhibition \cite{sczepanski2014cross}, we propose the reaction scheme
below, which is equivalent to a modification of Lente's reaction
scheme~\cite {lente2004homogeneous} through the additional process
representing the recycling of enantiomers:
\begin{align}
    &A+\R\xrightarrow{\;k_a\;} 2\R, && A+\L\xrightarrow{\;k_a\;} 2\L,\notag\\
    &A\xrightleftharpoons[k_d]{\,k_n\,} \R, && A\xrightleftharpoons[k_d]{\,k_n\,} \L.
    \label{rx:main}
\end{align}
Compared to Frank\rq{}s  model, the chiral inhibition is replaced by
linear decay reactions which model both recycling and non-autocatalytic
production. The rate constants are denoted by $k$, with the subscript
serving to identify the particular reaction. The only deterministic
fixed point of this model is the racemic state (\fig {stable}). This
model can be interpreted as a model of the evolution of early life
where primitive chiral self-replicators can be produced randomly
through non-autocatalytic processes at very low rates; the
self-replication is modeled by autocatalysis while the decay reaction
is a model for the death process.

We now approximate reaction scheme~\rx{main} by means of a
stochastic differential equation for the time evolution of the
chiral order parameter, $\omega$, which shows that in the regime
where autocatalysis is the dominant reaction, the functional form of
the multiplicative intrinsic noise from autocatalytic reactions
stabilizes the homochiral states. We consider a well-mixed system of
volume $V$ and total number of molecules $N$. As shown in the
Supplementary Material (SM), for $N \gg 1$, we obtain the following
equation for $\omega$, defined in the It\=o sense~\cite
{Gardiner2009}:
\begin{equation}
    \der{\omega}{t} = -\frac{2k_nk_d V}{Nk_a} \omega + \sqrt{\frac{2k_d}{N} (1-\omega^2)} \eta(t),
    \label{eq:langevin}
\end{equation}
where $\eta(t)$ is normalized Gaussian white noise~\cite{Gardiner2009}.


The time-dependent distribution of Equation~\eqref{eq:langevin} can be computed
exactly~\cite{biancalani2014noise,biancalani2015statistics}. The stationary
distribution~\cite{Gardiner2009},
\begin{equation}
    P_s(\omega) = \mathcal{N} \left(1-\omega^2\right)^{\alpha-1},\quad \text{with}\quad \alpha = \frac{Vk_n}{k_a},
    \label{eq:steady}
\end{equation}
depends on a single parameter, $\alpha$, where the normalization constant
$\mathcal{N}$ is given by
\begin{equation}
    \mathcal{N} = \left(\int_{-1}^{+1} \left(1-\omega^2\right)^{\alpha-1} d\omega\right)^{-1} = \frac{\Gamma\left(\alpha+\inv 2\right)}{\sqrt{\pi}\;\Gamma(\alpha)}.
\end{equation}

\begin{figure}[t]
    \hspace{-.2in}
    \includegraphics[width=.4\textwidth]{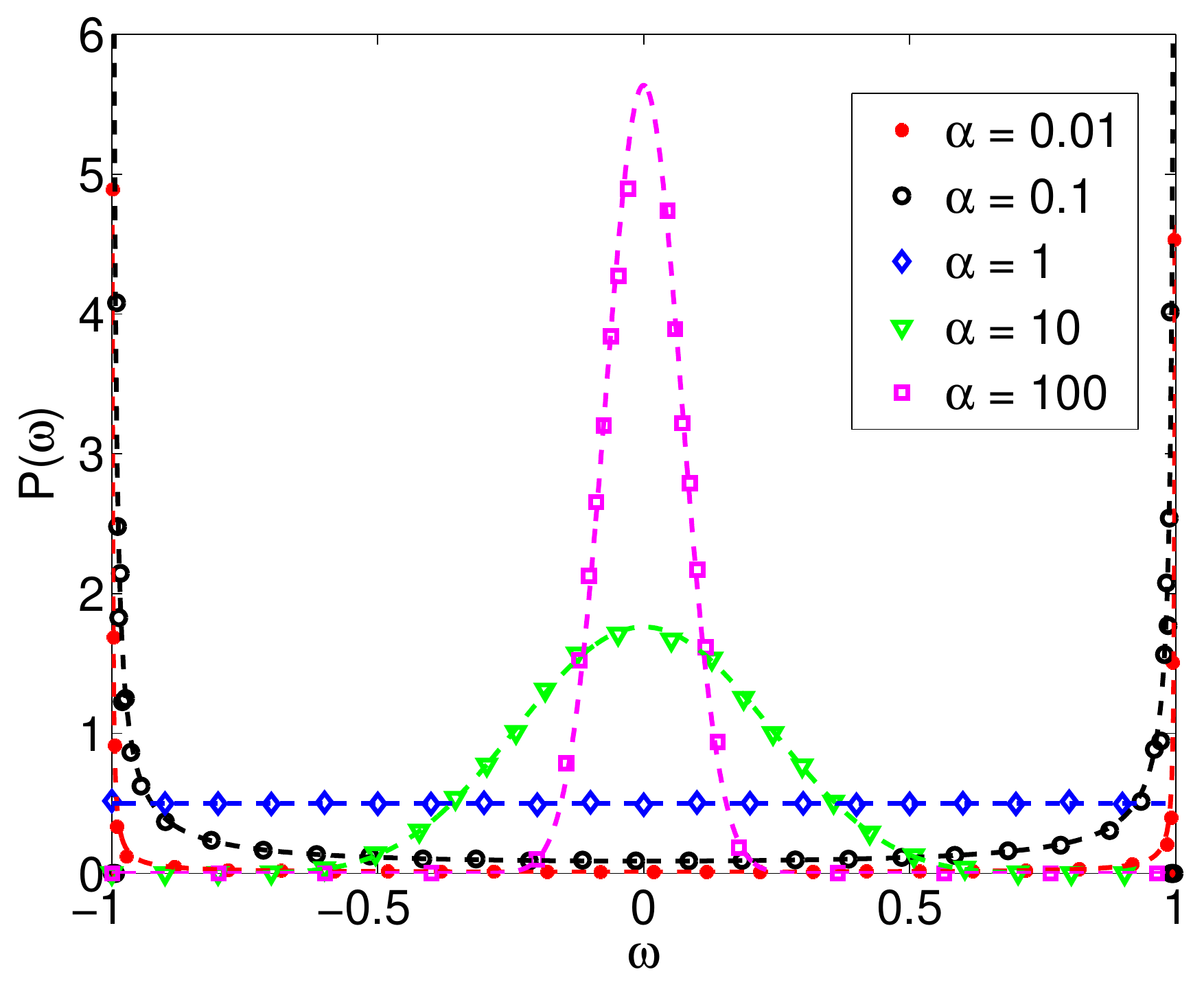}
    \caption{(Color online) Comparison between the stationary
    distribution, \eq{steady}, (dashed lines) and Gillespie
    simulations of reactions \rx{main}
    (markers), for different values of $\alpha$. Simulation
    parameters: $N = 10^3$, $k_a = k_n = k_d = 1$.}
    \vspace{-.2in}
    \label{fig:steady}
\end{figure}

Equation~\eqref{eq:steady} is compared in \fig{steady} against
Gillespie simulations~\cite{gillespie1977exact} of scheme \rx {main}.
For $\alpha = \alpha_c = 1$, $\omega$ is uniformly distributed. For
$\alpha \gg \alpha_c$, where the non-autocatalytic production is the
dominant production reaction, $P_{s}(\omega)$ is peaked around the
racemic state, $\omega = 0$. For $\alpha \ll \alpha_c$, where
autocatalysis is dominant, $P_{s}(\omega)$ is sharply peaked around the
homochiral states, $\omega = \pm 1$.  The simulations were performed
for $N=1000$, where the analytic theory is expected to be accurate; for
smaller values of $N$, the theory is qualitatively correct, but very small
quantitative deviations are observable compared to the simulations.
For example, for $N\sim 100$, $\alpha_c \sim 1.005$.



The deterministic part of \eq{langevin} has one fixed point at the
racemic state, consistently with the phase portrait in \fig{stable}.
The multiplicative noise in \eq{langevin} vanishes at homochiral
states, and admits its maximum at the racemic state. For
$\alpha\ll \alpha_c$, where autocatalysis is dominant, the amplitude
of the noise term in \eq{langevin} is much larger than the amplitude
of the corresponding deterministic term. In this regime, the system
ends up at homochiral states where the noise vanishes.

To understand this result physically, note that the source of the
multiplicative noise is the intrinsic stochasticity of the
autocatalytic reactions. While, on average, the two autocatalytic
reactions do not change the variable $\omega$, each time one of the
reactions takes place, the value of $\omega$ changes by a very small
discrete amount. As a result, over time the value of $\omega$ drifts
away from its initial value. Since the amplitude of the noise term
is maximum at racemic state and zero at homochiral states, this
drift stops at one of the homochiral states. The absence of the
noise from autocatalysis at homochiral states can be understood by
recognizing that at homochiral states, the molecules with only one
of two chiral states $\R$ and $\L$ are present, hence only the
autocatalytic reaction associated with that chiral state has a
non-zero rate. This reaction produces molecules of the same
chirality, keeping the system at the same homochiral state without
affecting the value of $\omega$, and therefore, the variable $\omega$
does not experience a drift away from the homochiral states due the
autocatalytic reactions.

Since the stationary distribution of $\omega$ in \eq{steady} is only
dependent on $\alpha$, the decay reaction rate, $k_d$ has no effect on
the steady state distribution of the system. The only role of this
reaction is to prevent the $A$ molecules from being completely
consumed, thus providing a well-defined non-equilibrium steady state
independent of the initial conditions. The parameter $\alpha$ is
proportional to the ratio of the non-autocatalytic production rate, $k_n$,
to the self-replication rate, $k_a$. In the evolution of early life,
when self-replication was a primitive function, $k_a$ would be small
and the value of $\alpha$ would therefore be large; but as
self-replication became more efficient, the value of $k_a$ would
increase and so $\alpha$ would decrease. Therefore, in our model, we
expect that life started in a racemic state, and it transitioned to
complete homochirality through the mechanism explained above, after
self-replication became efficient (i.e. when $\alpha \ll
\alpha_c$).

It is important to note that all of the previous mechanisms suggested for homochirality rely on assumptions that cannot be easily confirmed to hold during the emergence of life. However, even if all of such mechanisms fail during the origin of life, our mechanism guarantees the emergence of homochirality, since it only relies on self-replication and death, two processes that are inseparable from any living system.\\

\begin{figure}[b]
    \hspace{-.3in}
    \includegraphics[width=.45\textwidth]{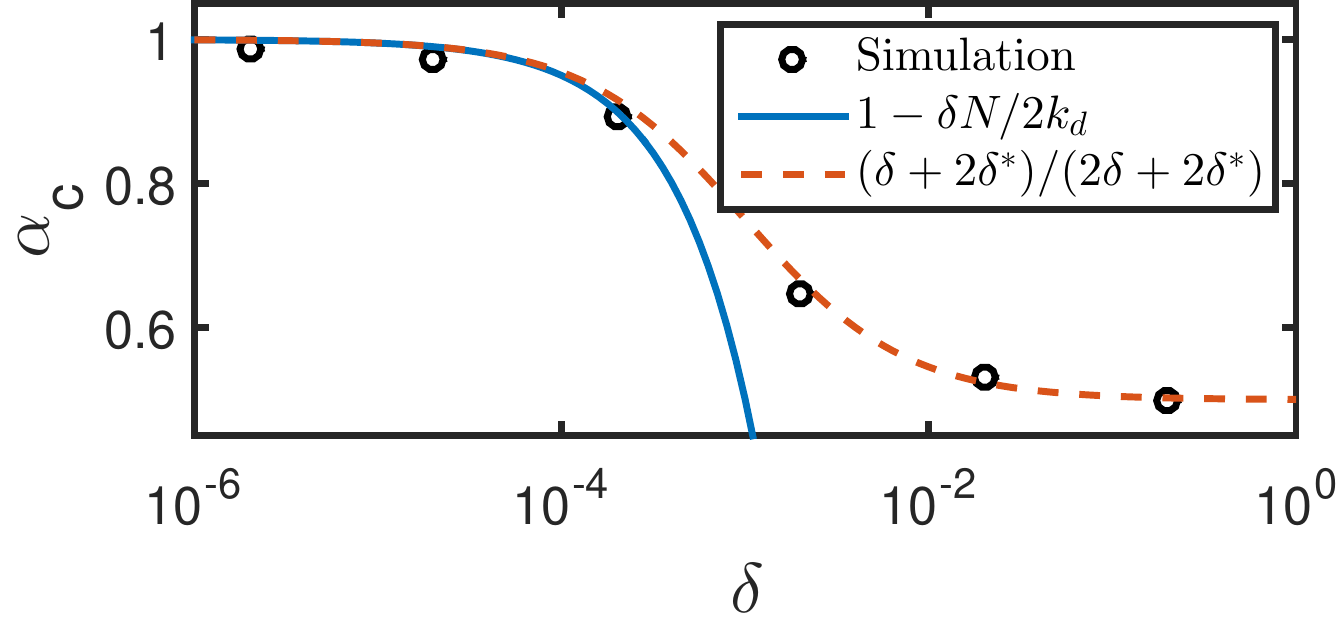}
    \caption{(Color online) Parameter $\alpha_{c}^{\text{patch}}$ in the
    two-patch system as a function of the diffusion rate $\delta$.
    Gillespie simulations (markers) are
    compared against \eq{smalld} (solid blue line) and \eq
    {interpolate} (dashed red line). Simulation parameters as in
    Fig.~\ref{fig:steady}.}
    \vspace{-.2in}
    \label{fig:twopatch}
\end{figure}
\begin{figure*}[t]
    \hspace{-.1in}
    \includegraphics[width=1\textwidth]{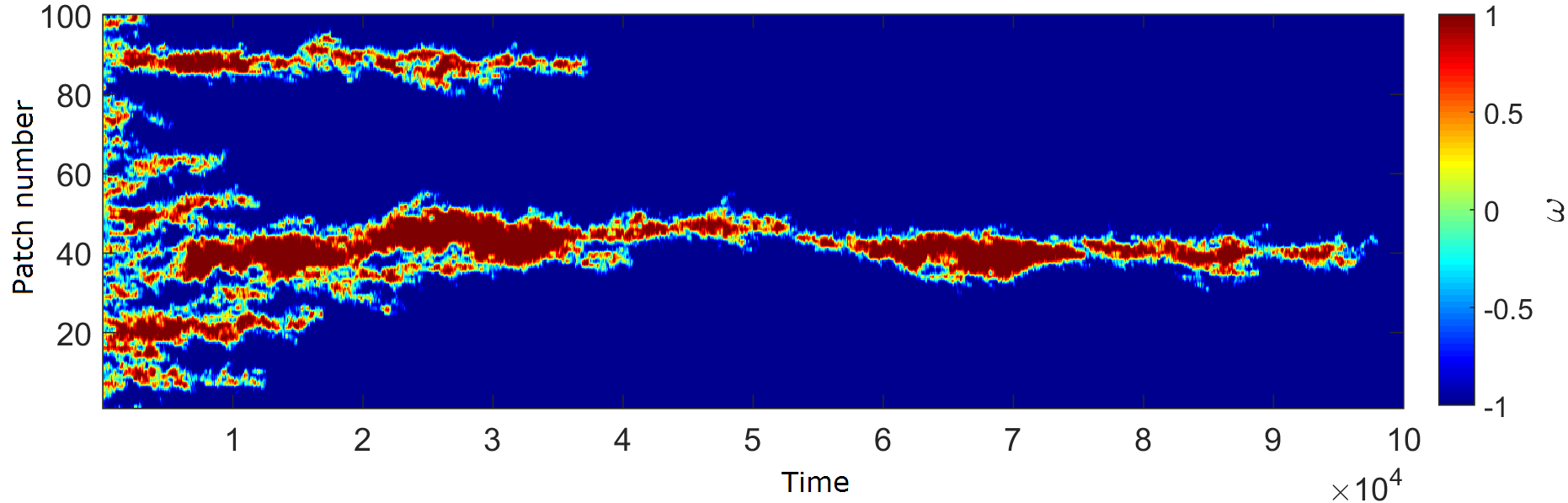}
    \caption{(Color online) Gillespie simulation of
    scheme \rx{spatial} for a one-dimensional system of $M = 100$
    patches, starting from racemic state and ending with all the
    patches in the same homochiral state $\omega = -1$. Simulation
    parameters: $N = 1000$, $k_a = k_d = 1$, $\delta = 10^{-3}$, and
    $k_n = 0$.}
    \vspace{-.2in}
    \label{fig:spatial}
\end{figure*}
\noindent {\it Stochastic model with spatial extension:-}
%
%
We now turn to the study of reaction scheme~\eqref{rx:main} generalized
to the spatially-extended case~\cite{mckane2004stochastic}. We
discretize space into a collection of $M$ patches of volume $V$,
indexed by $i$. The geometry of the space is defined by $\langle i
\rangle$ --- the set of patches that are nearest-neighbor to patch $i$
(e.g., for a linear chain, $\langle i \rangle = \{i-1,i+1\}$). We
indicate the molecules of species $A$ in patch $i$ by $A_i$ and
similarly for the other species. Each patch is well-mixed and
reactions~\eqref{rx:main} occur within, while molecules can diffuse
between neighboring patches with diffusion rate $\delta$. In summary, the following set of reactions
defines the spatial model:
\begin{equation} \label{rx:spatial}
\begin{split}
    &A_i\xrightleftharpoons[k_d]{\,k_n\,} \R_i,\quad A_i\xrightleftharpoons[k_d]{\,k_n\,} \L_i,\quad i=1,\ldots, M\\
    &A_i+\R_i\xrightarrow{\;k_a\;} 2\R_i,\quad A_i+\L_i\xrightarrow{\;k_a\;} 2\L_i\\
    &\R_i\xrightleftharpoons[]{\;\delta\;} \R_j, \quad \L_i\xrightleftharpoons[]{\;\delta\;} \L_j,\quad j\in\langle i\rangle.
\end{split}
\end{equation}
We now derive the following set of coupled stochastic differential
equation for the time evolution of the chiral order parameter
$\omega_i$, of each patch $i$ (see SM)
\begin{equation}
\begin{split}
    \der{\omega_i}{t} =& -\frac{2k_nk_d V}{Nk_a} \omega_i + \delta\sum_{j\in\langle i\rangle} (\omega_j-\omega_i)\\
    &+\sqrt{\frac{2k_d}{N}(1-\omega_i^2)}\eta_i(t) +
    \sqrt{\frac{\delta}{N}}\xi_i(\vec \omega, t),
    \label{eq:spatial}
\end{split}
\end{equation}
where now $N$ represents the average number of molecules per patch,
$\eta_i$\rq{}s are independent normalized Gaussian white
noises, $\xi_i$\rq{}s are zero mean Gaussian noise with correlator
\begin{equation}
\begin{split}
    \langle\xi_i(t)\xi_j(t\rq{})\rangle = &\left( 2\sum_{k\in\langle i\rangle}\left(1-\omega_i\omega_k\right) \delta_{i,j}\right .\\
    &\;\; \left.+ \left(\omega_i^2+\omega_j^2-2\right) \Chi_{\langle i\rangle}(j)\vphantom{\sum_{k\in\langle i\rangle}} \right) \delta(t-t\rq{}),
\end{split}
\end{equation}
and $\Chi_{\langle i\rangle}(j)$ is equal to one if $j\in\langle i\rangle$ and zero otherwise.

In order to see how the coupling of well-mixed patches affects their
approach to homochirality, it is instructive to consider the simplest
case of two adjacent patches ($M = 2$). In the two-patch model, various
scenarios can happen: the system may not exhibit homochirality
($\omega_1 \sim \omega_2 \sim 0$); each patch can separately reach
homochirality ($\omega_1 = \pm 1$ and $\omega_2 = \pm 1$); the system
exbihits global homochirality ($\omega_1 = \omega_2 = \pm 1$). We
first analyze the condition for each patch reaching homochirality using
perturbation theory, in the case of slow diffusion. The stationary
probability density function of the chiral order parameter of a single
patch, $Q_s(\omega)$ is defined by
\begin{equation} \label{eq:Q}
    Q_s(\omega) = \int_{-1}^{+1} Q_s(\omega,\omega_2) d\omega_2 = \int_{-1}^{+1} Q_s(\omega_1,\omega) d\omega_1,
\end{equation}
where $Q_s(\omega_1,\omega_2)$ is the joint probability distribution of
$\omega_1$ and $\omega_2$ at steady state from \eq{spatial}. If $\delta \sim k_d/N$ or
smaller, then (see SM) the stationary distribution
reads
\begin{equation}
    Q_s(\omega) = \mathcal Z (1-\omega^2)^{\alpha+\frac{\delta N}{2 k_d}-1},
\end{equation}
where $\mathcal Z$ is a normalization constant. This result shows
that the critical $\alpha$ in a single patch, up to the first order
correction in $\delta$, is given by
\begin{equation}
    \alpha_c^{\text{patch}} \approx 1 - \delta \frac{N}{2 k_d}, \quad
    \text{for }\delta \approx 0.
    \label{eq:smalld}
\end{equation}

We can now turn to the case of high diffusion. Recall that the patches
are defined as the maximum volume around a point in space in which the
system can be considered well-mixed. This can be interpreted as the
maximum volume in which diffusion dominates over the other terms acting
on the variable of interest (in this case $\omega$). From \eq{spatial},
this condition is fulfilled for $\delta\sim 2k_d \alpha/N$. In the
vicinity of the transition $\alpha$ is in order of one, therefore the
condition becomes $\delta \sim k_d/N$. For $\delta \gg k_d/N$, the
whole system can be considered well-mixed, and we can find the critical
value of $\alpha$ for each patch, starting from $\alpha_{c} = 1$, from
the well-mixed results, and using as volume the volume the whole
system, i.e., $MV$. This indicates that in a single patch
\begin{equation}
    \alpha_c^{\text{patch}} \approx \inv M, \quad \text{for }\delta \gg 0.
\end{equation}
A simple formula that interpolates between these extreme limits, asymptotic to
$1/M$ (with $M=2$) for large $\delta$ and to \eq{smalld} for small $\delta$, is
\begin{equation}
    \alpha_c^{\text{patch}} = \frac{\delta+2\delta^*}{2 \delta + 2\delta^*}, \hspace{.2in}\delta^* = \frac{k_d}{N}.
    \label{eq:interpolate}
\end{equation}

\Fig{twopatch} shows agreement between $\alpha_c^{\text{patch}}$
measured from Gillespie simulations of the two-patch system, and the
Eq.~\eqref{eq:interpolate}. At the parameter regime below the
$\alpha_c$ curve in~\fig{twopatch}, individual patches are homochiral.
Also, we find that the correlation between the homochiral states of the
two patches increases with diffusion rate $\delta$ and become
completely correlated when $\delta\sim k_d/N$ or more. In this regime
the system reaches global homochirality.

This latter result suggests that in the spatially-extended model, when
autocatalysis is the dominant reaction (i.e. $\alpha$ is small enough)
and when the diffusion rate is in the order of $k_d/N$ or larger, all
patches converge to the same homochiral state. \Fig {spatial} shows the
dynamics of a Gillespie simulation of a one-dimensional chain of $100$
patches, initializes at the racemic state, in the pure autocatalytic
limit ($k_n \to 0$). Very quickly, small islands of different
homochirality (blue and red) are formed. Islands of opposite chirality
competes against each other, until the system reaches global
homochirality. Note that for $\delta\sim k_d/N$ we can treat the
diffusion process deterministically by ignoring the last term in
\eq{spatial}. In this regime, \eq {spatial} is the same as the equation
describing one-dimensional voter model, implying that the transition to
homochirality is in the universality class of compact directed
percolation \cite {dickman1995hyperscaling}.

In conclusion, a racemic population of self-replicating chiral
molecules far from equilibrium, even in the absence of other
nonlinearities that have previously been invoked, such as chiral
inhibition, transitions to complete homochirality when the efficiency
of self-replication exceeds a certain threshold. This transition occurs
due to the drift of the chiral order parameter under the influence of
the intrinsic stochasticity of the autocatalytic reactions. The
functional form of the multiplicative intrinsic noise from
autocatalysis directs this drift toward one of the homochiral states.
Unlike some other mechanisms in the literature, this process does not
require an initial enantiomeric excess. In our model, the homochiral
states are not deterministic dynamical fixed points, but are instead stabilized by
intrinsic noise. Moreover, in the spatial extension of our model, we
have shown that diffusively coupled autocatalytic systems synchronize
their final homochiral states, allowing a system solely driven by
autocatalysis to reach global homochirality. We conclude that
autocatalysis alone is a viable mechanism for homochirality, without
the necessity of imposing chiral inhibition or other nonlinearities.

\begin{acknowledgments}
T.B. acknowledges valuable discussions with Elbert 
Branscomb. This material is based upon work supported by the 
National Aeronautics and Space Administration through the NASA 
Astrobiology Institute under Cooperative Agreement No. NNA13AA91A 
issued through the Science Mission Directorate.
\end{acknowledgments}
\bibliographystyle{apsrev4-1}
\bibliography{homochirality_ref}

\newpage
\clearpage
\twocolumngrid
  
\section{SUPPLEMENTARY MATERIAL}
\subsection{Well-Mixed System}
We start from reaction scheme (2) of the main paper
\begin{smequation} \label{rx:sm_main}
	\begin{split}
	&A+\R\xrightarrow{\;k_a\;} 2\R, A+\L\xrightarrow{\;k_a\;} 2\L,\\
	&A\xrightleftharpoons[k_d]{\,k_n\,} \R, A\xrightleftharpoons[k_d]{\,k_n\,} \L.
	\end{split}
\end{smequation}%
Each reaction changes the system from a state $\vec x \equiv (x_1,x_2,x_3) \equiv (a, d, l)$, specified by the concentration of molecules $A$, $\R$, and $\L$,  to a state of the form $\vec x + V^{-1} \vec s_m$ (for some $m\in\{1,\ldots, 4\}$), where $\vec s_m$ is the $m$\rq{}th row of the stoichiometry matrix
\begin{smequation}
	\mx S = \left(\begin{array}{ccc}
		-1	&	1	& 	0\\
		-1	&	0	&	1\\
		1	&	-1	&	0\\
		1	&	0	&	-1
		\end{array}\right).
\end{smequation}%
The probability per unit time of such transition is given by transition rates $T(\vec x + V^{-1} \vec s_m | \vec x)$ obtained from law of mass action for reactions \rx{sm_main}:
\begin{smequation} \label{eq:sm_transition}
\begin{split}
	T(\vec x + \e \vec s_1|\vec x) &= V (k_n + k_a d) a,\quad T(\vec x + \e \vec s_3|\vec x) =  V k_d d, \\
	T(\vec x + \e \vec s_2|\vec x) &=  V (k_n + k_a l) a,\quad T(\vec x + \e \vec s_4|\vec x) =V k_d l.
\end{split}
\end{smequation}%
The set of rates~\eqref{eq:sm_transition} is used to write the master equation for the time evolution of the probability density function, $P(\vec x, t)$, of the system being in the state $\vec x$ at time $t$~\cite{Kampen2007}. We begin by defining the functions $F_m$\rq{}s as
\begin{smequation}
	F_m(\vec x,t) = T(\vec x|\vec x-\e \vec s_m)P(\vec x-\e \vec s_m,t),
\end{smequation}
so that the master equation can be written as:
\begin{smequation} \label{eq:sm_master}
\begin{split}
	\pder{P(\vec x,t)}{t} = - \sum_{m=1}^{4} \left( F_m(\vec x+\e \vec s_m,t) - F_m(\vec x,t) \right).
\end{split}
\end{smequation}%
This equation defines the stochastic model and can be numerically simulated using the Gillespie algorithm~\cite{gillespie1977exact}.

In order to initiate an analytical treatment, we begin by expanding the right-hand side of the master equation (we follow~\cite{McKane2013}). We obtain a Gaussian noise approximation by truncating the expansion at the second-order, thus neglecting terms corresponding to higher moments. We arrive at the non-linear Fokker-Planck equation:
\begin{smequation} \label{eq:sm_fokker}
\begin{split}
	\pder{P}{t} \approx -\sum_{j=1}^{3}\pder{\left (H_j P \right)}{x_j} +\frac{1}{2}\sum_{j,k=1}^3\frac{\partial^2 \left( B_{jk} P \right) }{\partial x_j\partial x_k},
\end{split}
\end{smequation}%
where the drift vector $\vec H$ with component $H_j$ reads:
\begin{smequation}
\begin{split}
	\vec H =& \inv V \sum_m T(\vec x + \inv V \vec s_m)|\vec x) \vec s_m \\
		=& \left(
		\begin{array}{c}
			k_d(d+l)-a(2k_n+k_a(d+l))\\
			-k_d d+a(k_n+k_a d)\\
			-k_d l+a(k_n+k_a l)
		\end{array}\right).
\end{split}
\end{smequation}%
The symmetric diffusion matrix has the form
\begin{widetext}
\begin{smequation}
	\mx B = \inv{V^2}\sum_m T(\vec x + \inv V \vec s_m)|\vec x) \left(\vec s_m\otimes\vec s_m \right)
		= \inv V \left(
		\begin{array}{ccc}
			k_d(d+l)+a(2k_n+k_a(d+l))	&	-k_d d-a(k_n+k_a d)		& -k_d l-a(k_n+k_a l) \\
			-k_d d-a(k_n+k_a d) 		& 	k_d d+a(k_n+k_a d)		& 0\\
			-k_d l-a(k_n+k_a l)			&	0						& k_d l+a(k_n+k_a l)
		\end{array}
		\right),
\end{smequation}%
\end{widetext}
where the symbol $\otimes$ indicates the Kronecker product. We now decompose the diffusion matrix to $\mx B = \mx G \mx G^{T}$. Multiple choices for $\mx G$ exist~\cite{McKane2013}, and it is easy to check that the following $3\times 2$ matrix satisfies the decomposition:
\begin{smequation} \nonumber
\begin{split}
	\mx G = \inv {\sqrt{V}} \left(
		\begin{array}{cc}
			\sqrt{a \left(k_a d+ k_n\right)+ k_d d}		&	\sqrt{a (k_a l+ k_n)+ k_d l}\\
			-\sqrt{a \left(k_a d+ k_n\right)+ k_d d} 	& 	0\\
			0								&	-\sqrt{a (k_a l+ k_n)+ k_d l}	
		\end{array}\right).
\end{split}
\end{smequation}%
Equation~\eqref{eq:sm_fokker} is equivalent to the following stochastic differential equation (defined in the It\={o} sense)~\cite{Gardiner2009}
\begin{smequation} \label{eq:sm_lang}
	\der{\vec x}{t} = \vec H(\vec x) +\mx G(\vec x)\vec \eta(t)
\end{smequation}%
where $\eta_k$\rq{}s ($k=1,2$) are Gaussian white noises with zero mean and correlation 
\begin{smequation}
	\langle \eta_{j}(t) \eta_{k}(t') \rangle = \delta_{jk} \delta(t-t').
\end{smequation}%
Note that since, the Fokker-Planck equation (\ref{eq:sm_fokker}) only depends on $\mx B$ and not the particular choice of its decomposition $\mx G$, the probability density function of $\vec x$ and its time evolution do not depend on $\mx G$ either~\cite{McKane2013}.

The number of degrees of freedom in \eq{sm_lang} can be reduced by noting two facts: (i) the reaction scheme \rx{sm_main} conserves the total number of molecules, meaning that the total concentration $n = a+d+l$ is conserved; (ii) simulations show that the concentration $r=d+l$ settles to a Gaussian distribution around its fixed point value $r^*$, allowing us to substitute $r(t) \to r^*$. We therefore change variables in Eq.~\eqref{eq:sm_lang} (using It\=o\rq{}s formula)~\cite{Gardiner2009}, from 
\begin{smequation}
	\left(\begin{array}{c}
		a\\
		d\\
		l
		\end{array}\right) \to \left(\begin{array}{c}
		n\\
		r\\
		\omega
		\end{array}\right) = \left(\begin{array}{c}
		a+d+l\\
		d+l\\
		(d-l)/(d+l)
		\end{array}\right),
\end{smequation}%
so that the only dynamics occurs in the chiral order parameter $\omega$. In the new variables, we find that $\dot n = 0$ and, by taking the positive solution of $\dot r = 0$, that is
\begin{smequation}
	r^{*} = \frac{\sqrt{(k_{a} n-k_{d}-2 k_{n})^2+8 k_{a} k_{n} n}+k_{a} n-k_{d}-2 k_{n}}{2 k_{a}},
\end{smequation}%
we substitute $r\to r^{*}$ in the equation for $\omega$, and use the rule for summing Gaussian variables (i.e. $a \eta_{1} + b \eta_{2} = \sqrt{a^{2}+b^{2}} \eta$; where $a$ and $b$ are generic functions~\cite{Gardiner2009}) to express the stochastic part of the equation using a single noise variable. Expressing the result in terms of the total number of molecules $N = V n$, for $N \gg 1$, we arrive at the following stochastic differential equation for chirality order parameter $\omega$ (equation (2) of the main text):
\begin{smequation}
	\der{\omega}{t} = -\frac{2k_nk_d V}{Nk_a} \omega + \sqrt{\frac{2k_d}{N} (1-\omega^2)} \eta(t),
	\label{eq:sm_langevin}
\end{smequation}%
where $\eta(t)$ is Gaussian white noise with zero mean and unit variance. The corresponding Fokker-Planck equation of \eq{sm_langevin} is an exactly solvable partial differential equation with time dependent solution given in~\cite{biancalani2015statistics}. The steady state probability distribution of $\omega$ is given by
\begin{smequation}
	P_s(\omega) = \mathcal{N} \left(1-\omega^2\right)^{\alpha-1},\quad \text{with}\quad \alpha = \frac{Vk_n}{k_a},
	\label{eq:sm_steady}
\end{smequation}%
with the normalization constant
\begin{smequation}
	\mathcal{N} = \left(\int_{-1}^{+1} \left(1-\omega^2\right)^{\alpha-1} \text{d}\omega\right)^{-1} = \frac{\Gamma\left(\alpha+\inv 2\right)}{\sqrt{\pi}\;\Gamma(\alpha)}.
\end{smequation}%

\subsection{Two-patch model}
Starting from reaction scheme (6) of the main paper for the spatial extension of our model 
\begin{smequation} \label{rx:sm_spatial}
\begin{split}
	&A_i\xrightleftharpoons[k_d]{\,k_n\,} \R_i,\quad A_i\xrightleftharpoons[k_d]{\,k_n\,} \L_i,\quad i=1,\ldots, M\\
	&A_i+\R_i\xrightarrow{\;k_a\;} 2\R_i,\quad A_i+\L_i\xrightarrow{\;k_a\;} 2\L_i\\
	&\R_i\xrightleftharpoons[]{\;\delta\;} \R_j, \quad \L_i\xrightleftharpoons[]{\;\delta\;} \L_j,\quad j\in\langle i\rangle,
\end{split}
\end{smequation}%
for $M = 2$, we can follow the procedure explained in the previous section to obtain a Fokker-Planck equation for time evolution of the probability density of system being at a state with concentrations $a_1$, $d_1$, $l_1$, $a_2$, $d_2$, and $l_2$. Again we can reduce the number of variables using the following facts (i) the total concentration $n_t = n_1+n_2 = a_1+d_1+l_1+a_2+d_2+l_2$ is conserved; (ii) simulation shows that in long time, the variables $r_1 = d_1 + l_1$, $r_2 = d_2 + l_2$, and $\Delta = n_1-n_2$ settle to Gaussian distributions around their fixed point values $r_1 = r_2 = r^*$ and $\Delta = 0$. We do the following change of variables 
\begin{smequation}
	\left(\begin{array}{c}
			a_1\\d_1\\l_1\\a_2\\d_2\\l_2
		\end{array}\right) \to \left(\begin{array}{c}
			n_t\\\Delta\\r_1\\r_2\\\omega_1\\\omega_2
		\end{array}\right) = \left(\begin{array}{c}
			a_1+d_1+l_1+a_2+d_2+l_2\\
			a_1+d_1+l_1-a_2-d_2-l_2\\
			d_1+l_1\\
			d_2+l_2\\
			(d_1-l_1)/(d_1+l_1)\\
			(d_2-l_2)/(d_2+l_2)
		\end{array}\right)
\end{smequation}%
 using It\=o\rq{}s formula. Now the dynamics only occurs only in $\vec \omega = (\omega_1, \omega_2)$. For large average number of molecules per patch $N\gg1$, the resulting Fokker-Planck equation for time evolution of the joint probability density function of $\omega_1$ and $\omega_2$, $Q(\vec\omega,t)$, reads
\begin{smequation} \label{eq:sm_fokker2}
\begin{split}
	\pder{Q}{t} = -\sum_{i=1}^{2}\pder{\left (\left(\mx L\vec \omega\right)_i Q \right)}{\omega_i} +\frac{1}{2}\sum_{i,j=1}^2\frac{\partial^2 \left( U_{ij} Q \right) }{\partial \omega_i\partial \omega_j}.
\end{split}
\end{smequation}%
Note that the above sums are now over the patches, and not over species as in Eq.~\eqref{eq:sm_fokker}. The Jacobian matrix $\mx L$ 
\begin{smequation}
	\mx L = -\frac{2k_d k_n V}{N k_a} \left(\begin{array}{cc}
			1	&	0\\
			0	& 1
		\end{array} \right) + \delta\left(\begin{array}{cc}
			-1	&	1\\
			1	& 	-1
		\end{array}\right), \label{eq:A}
\end{smequation}%
and the diffusion matrix $\mx U$ reads
\begin{smequation} \label{eq:U}
\begin{split}
	\mx U = &\;\frac{2k_d}{N}\left(\begin{array}{cc}
			1-\omega _1^2	&	0\\
			0			&	1-\omega _2^2
		\end{array}\right)\\&+\frac{\delta}{N}\left(\begin{array}{cc}
			2(1-\omega _1\omega_2)		&	\omega_1^2+\omega_2^2-2\\
			\omega_1^2+\omega_2^2-2	&	2(1-\omega _1\omega_2).
		\end{array}\right).
\end{split}
\end{smequation}%
Note that the stochastic differential equation corresponding to \eq{sm_fokker2} is the $M=2$ case of equation (7) of the main paper.

\subsection{Perturbation theory for the small diffusion case}
Now, we can use perturbation theory for small $\delta$, to find the stationary probability density of $\omega$ for each patch defined as
\begin{smequation} \label{eq:sm_Q}
	Q_s(\omega) = \int_{-1}^{+1} Q_s(\omega,\omega_2) \text{d}\omega_2 = \int_{-1}^{+1} Q_s(\omega_1,\omega) \text{d}\omega_1,
\end{smequation}%
where $Q_s(\omega_1,\omega_2)$ is the stationary solution of \eq{sm_fokker2}. For $\delta \sim k_d/N$ or smaller, we can treat the diffusion deterministically by ignoring the last term in \eq{U}.
To solve for $Q_s(\omega)$, we begin by rewriting Eq.~\eqref{eq:sm_fokker2} as a continuity equation,
\begin{smequation}
	\partial_t Q + \nabla \cdot \vec J = 0,
\end{smequation}%
which defines the probability current $\vec J$ as~\cite{Gardiner2009}
\begin{smequation}
	\vec J = \mx L \vec \omega \;Q - \inv 2 \nabla \cdot \left(\mx U Q\right).
\end{smequation}%
By the conservation of probability, at stationary conditions, the total probability flux $\vec J_s$ through each vertical section of $\omega_1$-$\omega_2$ plane must be zero. That is
\begin{smequation} \label{eq:sm_flux}
\begin{split}
	&\int_{-1}^{+1} J_{s,1} \text{d}\omega_2 =\int_{-1}^{+1} \left((\mx L \vec \omega)_1 Q_s-\inv 2\partial_{\omega_1}(U_{11}Q_s)\right) \text{d}\omega_2 \\
		&=\;Q_s(\omega_1)\omega_1\left(\frac{2k_d}{N}(1-\alpha)-\delta\right)-\frac{k_d}{N}(1-\omega_1^2)\der{Q_s}{\omega_1}\\
		&+\delta \int_{-1}^{+1}\omega_2 Q_s(\omega_1,\omega_2)\text{d}\omega_2 = 0.
\end{split}
\end{smequation}%
The last integral can be evaluated using Bayes' theorem
\begin{smequation}
\begin{split}
	\delta \int_{-1}^{+1}\omega_2&Q_s(\omega_1,\omega_2)\text{d}\omega_2 =\delta\int_{-1}^{+1}\omega_2Q_s(\omega_2|\omega_1)Q_s(\omega_1)\text{d}\omega_2\\
		&= \delta\; Q_s(\omega_1)\langle \omega_2\rangle_{\omega_1} = \mathcal{O}(\delta^2),
\end{split}
\end{smequation}%
which is of order $\delta^2$ for small $\delta$, since, $\langle \omega_2 \rangle_{\omega_1}$ (the expected value of $\omega_2$ given $\omega_1$) vanishes at zero $\delta$, and therefore, of order $\delta$ for small $\delta$. In this regime, \eq{sm_flux} provide us with a differential equation for $Q_s(\omega)$ with the solution (equation (10) of the main paper)
\begin{smequation}
	Q_s(\omega) = \mathcal Z (1-\omega^2)^{\alpha+\frac{\delta N}{2 k_d}-1},
\end{smequation}%
where the normalization constant $\mathcal Z$ is given by
\begin{smequation}
	\mathcal{Z} = \frac{\Gamma\left(\alpha+\frac{\delta N}{2 k_d}+\inv 2\right)}{\sqrt{\pi}\;\Gamma(\alpha+\frac{\delta N}{2 k_d})}.
\end{smequation}


\end{document}